# Compression as a universal principle of animal behavior

Short title: Compression in animal behavior

Ramon Ferrer-i-Cancho[1,]*, Antoni Hernández-Fernández[1,2], David Lusseau[3,4], Govindasamy Agoramoorthy[5], & Minna J. Hsu[6], Stuart Semple[7]

(1) Complexity and Quantitative Linguistics Lab. Departament de Llenguatges i Sistemes Informàtics. TALP Research Center. Universitat Politècnica de Catalunya, Barcelona (Catalonia), Spain.

(2) Departament de Lingüística General. Universitat de Barcelona, Barcelona (Catalonia), Spain.

(3) Institute of Biological and Environmental Sciences, University of Aberdeen, Aberdeen, UK.

(4) Marine Alliance Science and Technology for Scotland, University of Aberdeen, Aberdeen, UK.

(5) College of Environmental and Health Sciences, Tajen University, Yanpu, Taiwan, ROC.

(6) Department of Biological Sciences, National Sun Yat-sen University, Kaohsiung, Taiwan, ROC.

(7) Centre for Research in Evolutionary and Environmental Anthropology, University of Roehampton, London, UK.

*Author for correspondence (rferrericancho@lsi.upc.edu).

Keywords: law of brevity, compression, animal communication, language, animal behavior

ABSTRACT

A key aim in biology and psychology is to identify fundamental principles underpinning the behavior of animals, including humans. Analyses of human language and the behavior of a range of non-human animal species have provided evidence for a common pattern underlying diverse behavioral phenomena: words follow Zipf's law of brevity (the tendency of more frequently used words to be shorter), and conformity to this general pattern has been seen in the behavior of a number of other animals. It has been argued that the presence of this law is a sign of efficient coding in the information theoretic sense. However, no strong direct connection has been demonstrated between the law and compression, the information theoretic principle of minimizing the expected length of a code. Here we show that minimizing the expected code length implies that the length of a word cannot increase as its frequency increases. Furthermore, we show that the mean code length or duration is significantly small in human language, and also in the behavior of other species in all cases where agreement with the law of brevity has been found. We argue that compression is a general principle of animal behavior, that reflects selection for efficiency of coding.

## 1. INTRODUCTION

Understanding the fundamental principles underpinning behavior is a key goal in biology and psychology (Grafen, 2007; Gintis, 2007). From an evolutionary perspective, similar behavioral patterns seen across different animals (including humans) may have evolved from a common ancestral trait, or may reflect convergent evolution; looking for shared quantitative properties of behavior across diverse animal taxa may thus allow identification of the elementary processes constraining or shaping behavioral evolution (e.g., Hailman et al., 1985; McCowan et al. 2002; Sumpter, 2006). Recent evidence of consistent patterns linking human language with vocal communication and other behavior in a range of animal species has pointed to the existence of at least one such general principle (Ferrer-i-Cancho & Lusseau, 2009; Semple et al., 2010). Here, we detail the nature of this evidence, before describing and mathematically exploring the principle in question – compression – that we propose underlies the consistent patterns discovered.

Words follow Zipf's law of brevity, i.e. the tendency of more frequently used words to be shorter (Zipf, 1935, Strauss et al., 2007), which can be generalized as the tendency of more frequent elements to be shorter or smaller (Ferrer-i-Cancho & Hernández-Fernández, 2013). Statistical laws of language have been studied outside human language (see Ferrer-i-Cancho & Hernández-Fernández, 2013 for an overview), and to our knowledge, the first report of conformity to this generalized brevity law outside human language is found in the pioneering quantitative research by Hailman et al. (1985) on chick-a-dee calls. More recently, accordance with the generalized law of brevity has been found in dolphin surface behavioral patterns (Ferrer-i-Cancho & Lusseau, 2009), the vocalizations of Formosan macaques (Semple et al., 2010) and a subset of the vocal repertoire of common marmosets (Ferrer-i-Cancho & Hernández-Fernández, 2013). A lack of conformity to the law has been found in the vocalizations of golden-backed uakaris (Bezerra et al., 2011) and ravens (Ferrer-i-Cancho & Hernández-Fernández, 2013).

Statistical patterns of language, and the law of brevity in particular, offer a unique chance to reframe research on language universals (Evans & Levinson, 2009) in three ground-breaking directions, according to the current state of the art. Firstly, in extending typology of linguistic universals beyond human language and considering the possibility of universals of behavior across species. Secondly, in considering that some language universals may, in fact, not be specific to language or communication but an instance of universals of animal behavior. Thirdly, in changing the stress of the quest from universal properties of language to universal principles of language (or behavior) along the lines of modern quantitative linguistics (Zipf, 1949; Köhler, 1987; Köhler, 2005). More important than expanding the collection of universal properties or delimiting what is universal or not, is (a) a deep understanding of the principles explaining the recurrence of these regularities regardless of the context and (b) the role of those principles, perhaps hidden, in exceptions to widespread regularities. If these regularities are not inevitable (Ferrer-i-Cancho et al., 2013), a parsimonious hypothesis would be a minimal set of principles that are independent from the context and thus universal. An important research goal is defining such a set, if it exists. Furthermore, this third point could contribute to reconciling the wide diversity of languages with the need for unifying approaches, outside of the realm of a "language faculty" or innate specializations for language (Evans & Levinson, 2009).

The discovery of conformity to statistical laws of language outside human language raises a very important research question: are the findings simply a coincidence, or are there general principles responsible for the appearance of the same statistical pattern across species, or across many levels of life? As the arguments against the simplicity of language laws within and outside human language are falling (e.g., Ferrer-i-Cancho & Elvevåg, 2010; Ferrer-i-Cancho & McCowan, 2012; Hernández-Fernández et al., 2011; Ferrer-i-Cancho et al., 2013), here we propose the principle of compression - the information theoretic principle of minimizing the expected length of a code - as an explanation for the conformity to the law of brevity across species. Hereafter, we consider the term "brevity", the short length of an element, as an effect of "compression".

In his pioneering research, G. K. Zipf argued that the law of brevity was a consequence of a general principle of economy (Zipf, 1949). He used the metaphor of an "...artisan who will be obliged to survive by performing jobs as economically as possible with his tools" (p. 57). The artisan is a metaphor for a speaker (or a community of speakers) while tools is a metaphor for words. The artisan should decrease the size or the mass of the tools that he uses more frequently to reduce the amount of work (Zipf, 1949, pp. 60-61). Along similar lines, it is well-known that S. Morse and A. Vail optimized the Morse code simply by choosing the length of each character approximately inversely proportionally to its frequency of occurrence in English (Gleick, 2011). Although it has been claimed that conformity to the law of brevity is a sign of efficient coding in both human language (Zipf, 1949) and animal behavior (Ferrer-i-Cancho & Lusseau, 2009; Semple et al., 2010), no strong direct connection with standard information theory, and with coding theory in particular, has been shown.

Imagine that

- $n$ is defined as the number of elements of the repertoire or vocabulary.

- $p_i$ is defined as the probability of producing the $i$-th most likely element.

- $e_i$ is defined as the energetic cost of that element. $e_i$ could be the number of letters or syllables of the $i$-th most likely word or the mean duration of the $i$-th most likely vocalization or behavior of a non-human animal.

Then, the mean energetic cost of a repertoire or vocabulary can be defined as

$$E = \sum_{i=1}^{n} p_i e_i. \tag{1}$$

Indeed, a particular case of the definition of $E$ in Eq. 1 is $E_{CL}$, the mean code length, which is the cost function that is considered in standard information theory for the problem of compression, being $e_i$ the length of the code used to encode the $i$-th element of the set of symbols (Cover & Thomas, 2006, pp. 110). Data compression consists of finding the code lengths that minimize $E_{CL}$ given the probabilities $p_1,...p_i,...,p_n$. $E_{CL}$ provides an objective measure of coding efficiency. In his pioneering research, G. K. Zipf called his own version of Eq. 1 the "minimum equation" (in the sense that it is the function to minimize in order to reduce effort), with $p_i$ being frequency of use of a tool, and $e_i$ being work, i.e. the product of the "mass" of the tool and the "distance" of the tool to the artisan (Zipf, 1949, p. 59). Therefore, G. K. Zipf's view constitutes a precursor of coding theory. However, he never tested if $E_{CL}$ is significantly small in human languages, using a rigorous statistical approach. Hereafter, the definition of $E_{CL}$ from information theory is relaxed so that $e_i$ can be not only the bits used to code the $i$-th element but also its size, length or duration. $e_i > 0$ is assumed here as we focus on 'elements' that can be observed in a real system, although some models of behavior produce elements of length zero, and then $e_i = 0$ for such empty elements (see Ferrer-i-Cancho & Gavaldà 2009 for a review of these models). Pressure for minimizing $E_{CL}$ in animal behavior may arise from a number of different sources.

Firstly, there may be a need to minimize the direct energetic costs of producing a behavior; evidence from a range of vertebrate and invertebrate species indicates, for example, that energy availability may limit the duration of calling behavior (Gillooly & Ophir, 2012; Thomas, 2002; Klump, 2005; Bennet-Clark, 1998; Fletcher, 1997). Furthermore, Waters & Jones (1995) showed theoretically that energy is directly proportional to duration in acoustic communication by integrating the energy flux density, as it is well-known in fluid mechanics, according to which the sound energy flux (a) is given by the time integral of the squared sound pressure (Landau & Lifshitz, 1987) and (b) can be roughly approximated by $A \cdot t$, where $t$ is the pulse duration and $A$ is its amplitude, for relatively short stimuli, as Baszczyk (2003) explains. According to general acoustics, the energy of a sound wave is $\xi = P \cdot t$, where $P$ is its sound power and $t$ is its duration and $P = I \cdot S$, where $I$ is the sound intensity and $S$ is the area of the propagation surface (Kinsler et al., 2000). For a sound wave of amplitude $A$, the sound energy becomes (Fahy, 2001, Kinsler et al., 2000):

$$\xi = P{\cdot}t = I{\cdot}S{\cdot}t \propto A^2{\cdot}t \quad , \tag{2}$$

namely, the energy of a signal depends linearly on time and the squared amplitude. This suggests *a priori* it is not only duration that matters but also amplitude. However, amplitude determines the energy of a sound wave at a given time and therefore reducing the amplitude reduces the reach of the signal due to the natural degradation of the signal with distance, and interference caused by other sounds (Bennet-Clark, 1998). Amplitude fluctuations during the propagation of sound have different effects on receivers, depending on wave frequency, and interfere with their ability to detect directionality (Wiley & Richards, 1978). Minimizing energy by amplitude modulation puts the success of communication at risk. Furthermore, amplitude is highly determined by body size (Bennet-Clark, 1998; Fletcher, 2004) as Gillooly & Ophir (2010) demonstrate, because there is a dependency between sound power and amplitude. Therefore, our energy function $E_{CL}$ is capturing the contribution to sound energy that can be more easily controlled, namely that of duration.

Secondly, shorter signals may have advantages independent of energetic production costs. In predation contexts, it may be beneficial for calling prey to decrease conspicuousness to predators (Ryan et al., 1982; Endler, 1993; Hauser, 1996) or for calling predators to decrease conspicuousness to prey (Deecke et al., 2005). In addition, shorter signals suffer less from problems linked to reverberation, an important phenomenon that degrades signals in environments containing solid structures (Waser & Brown, 1986). For the vocalizations of rainforest primates, for example, there appears to be an upper call duration limit of 200-300ms, to minimize such interference (Brown & Sinott, 2006, p. 191). Interestingly in relation to this point, the law of brevity has been found in one subset of the repertoire of common marmosets where all but one call type (the exception being the submissive squeal) are below the 300ms limit; the other subset, where the law is not found, (a) consists of calls that are above 300ms in duration and (b) contains all long-distance communication calls (Ferrer-i-Cancho & Hernández-Fernández, 2013). If sound pressure, *P,* falls inversely proportional to the distance from the sound source (according to the so-called distance law of sound attenuation), and sound intensity (and then energy, recall $\xi = I \cdot S$) falls inversely proportional to the square of the distance (Landau & Lifshitz, 1987; Kinsler et al 2000) then long-distance calls must generally show an increase either in sound intensity or duration, or both. Therefore, $E_{CL}$ does not measure all the costs of a repertoire, as it does not include intensity and sometimes it may be advantageous to increase duration rather than reducing it. By focusing on $E_{CL}$ rather than on $E$, we hope to shed light on the importance of compression in animal behavior.

Here we aim to provide support for the principle of compression (minimization of $E_{CL}$) as a general principle of animal behavior. We do not propose that compression is the only principle of animal behavior, or the only principle by which behavior is optimized. The design of language is a multiple constraint engineering problem (Evans & Levinson, 2009; Köhler, 2005) and the same applies to the communication systems of other species (Endler, 1993). The appearance of design in communication systems can exist without a designer or intentional engineering (Cornish, 2010, Kirby et al., 2008).

Our notion of principle should not be confused with explanation. Principles are the ingredients of explanations. Using the physical force of gravity as an example helps to illustrate our notion of principle. The universal force of gravity explains why objects fall, but when a rocket flies upwards its movement does not constitute an exception to the force. The force is still acting and is involved in explaining, for instance, the amount of fuel that is needed to fly in the opposite direction of the force. As the falling of an object is a manifestation of the force of gravity, we propose the law of brevity in animal behavior is a manifestation of a principle of compression. Just as the force of gravity is still acting on a rocket moving away from the Earth, so we should not conclude prematurely that compression is not a relevant principle of behavior when the law of brevity is not detected.

Although the Earth and Venus are very different planets, the force of gravity is valid in both (indeed universal) and physicists only care about the difference in its magnitude. Similarly, we do not assume that human language and animal behavior cannot have compression in common, no matter how large the biological, social, cognitive or other differences are. Science is founded on parsimony (Occam's razor), and from an evolutionary perspective hypothesizing that the principle of compression is common to humans and other animal species is *a priori* simpler than hypothesizing that compression is not shared. We are adopting the perspective of standard statistical hypothesis testing, where the null hypothesis is that there is no difference between two populations e.g. two different species (Sokal and Rohlf, 1995). The research presented here strongly suggests that there is no need to adopt, *a priori*, the more demanding alternative hypothesis of an intrinsic difference between humans and other species' linguistic and non-linguistic behavior for the particular case of the dependency between the size or length the units and their frequency. However, according to modern model selection theory, a good model of reality has to comply with a trade-off between its parsimony and the quality of its fit to reality (Burnham & Anderson, 2004). A key and perhaps surprising result that will be presented in this article is that $E_{cl}$ is never found to be significantly high, in spite of apparently clear advantages in certain situations of increasing signal/behavior length (e.g. for long distance communication). It could interpreted that even when the direct effect of compression is not observed, compression still has a role, just as a rocket heading to space is still being attracted by the Earth's gravitational field. It is still too early to conclude that compression has no role even when there is no evidence that $E_{cl}$ is being minimized; it is important to note that there are serious statistical limits for detecting efficient coding, especially in small repertoires (Ferrer-i-Cancho & Hernández-Fernández, 2013).

The remainder of the article is organized as follows. Section 2 shows that the energetic cost of an element (e.g. the length of a word) cannot increase as frequency increases in a system that minimizes $E_{cl}$. Thus, the law of brevity is an epiphenomenon of compression. Section 3 shows direct evidence of compression in animal behavior, in particular by demonstrating that $E_{cl}$ is significantly small in all the cases where conformity to the law of brevity has previously been found. Section 4 discusses these findings.

## 2. THE MATHEMATICAL RELATIONSHIP BETWEEN COMPRESSION AND THE LAW OF BREVITY

By definition of $p_i$, one has

$$p_1 \geq p_2 \geq p_3 \geq \dots p_n. \qquad (3)$$

If Zipf's law of brevity was agreed with fully, one should also have

$$e_1 \leq e_2 \leq e_3 \leq \dots e_n. \qquad (4)$$

Here a mathematical argument for Zipf's law of brevity as a requirement of minimum cost communication is presented.

Imagine that the law of brevity is not agreed with fully, namely that there exist some $i$ and $j$ such that $i < j$, $e_i > e_j$, and thus Eq. 4 does not hold perfectly. One could swap the values of $e_i$ and $e_j$. That would have two important consequences. Firstly, the agreement with the law of brevity (Eq. 4) would increase. Secondly, $E_{CL}$ would decrease as is shown next. To see that $E_{CL}$ is reduced by swapping $e_i$ and $e_j$, we calculate $E^s_{CL}$, the mean cost after swapping,

$$E^s_{CL_s} = E_{CL} - p_i e_i - p_j e_j + p_i e_j + p_j e_i. \qquad (5)$$

After rearranging Eq. 5, it is obtained that the increment of cost, $\delta = E^s_{CL}$ - $E_{CL}$ is

$$\delta = (p_i - p_j)(e_j - e_i). \qquad (6)$$

$\delta < 0$ means that the cost has reduced i.e. compression has increased. To see that $\delta \leq 0$ with equality if and only if $p_i = p_j$, notice that $p_i - p_j \geq 0$ (by definition of $p_i$ and $p_j$ and $i < j$) and $e_j - e_i < 0$ (recall we are considering the case $e_i > e_j$ and thus $e_i = e_j$ is not possible). In sum, a system that minimizes $E_{CL}$ needs to follow Zipf's law of brevity (Eq. 4), otherwise there is a pair of element costs ($e_i$ and $e_j$) that can be swapped to reduce $E_{CL}$. Finally, note that the idea of swapping of costs was introduced to demonstrate the necessity of the law of brevity (Eq. 4) from the minimization of $E_{CL}$, not to argue that swapping is the evolutionary mechanism through which durations or length are optimized according to frequency.

## 3. DIRECT EVIDENCE OF COMPRESSION

### 3.1. Methods.

$E_{CL}$ is estimated taking $p_i$ as the relative frequency of the $i$-th most frequent element of a sample. If $E_{CL}$ is significantly small in animal behavior that would be a sign of compression to some degree. Whether $E_{CL}$ is significantly small or not can be determined by means of a randomization test (Sokal & Rohlf, 1995, pp. 803-819). To this end, $E_{CL}'$, a control $E_{CL}$, is defined as

$$E_{CL}' = \sum_{i=1}^{n} p_i e_{\pi(i)}. \tag{7}$$

where $\pi(i)$ is a permutation function (i.e. a one-to-one mapping from and to integers 1,2,...,$n$). The left p-value of the test is defined as the proportion of permutations where $E_{CL}' \leq E_{CL}$ and the right p-value is defined as the proportion of permutations where $E_{CL}' \geq E_{CL}$. The left p-value is estimated by $T_L/T$, where $T$ is the total number of random permutations used and $T_L$ is the number of such permutations where $E_{CL}' \leq E_{CL}$. Similarly, the right p-value is estimated by $T_R/T$, where $T_R$ is the number of random permutations where $E_{CL}' \geq E_{CL}$. $T=10^5$ uniformly random permutations were used.

### 3.2. Data

We adopt the term type for referring not only to word types, but also to types of vocalization and behavioral patterns. We used a dataset that comprises type frequency and type size/length/duration from the following species: dolphins (*Tursiops truncatus*), humans (*Homo sapiens sapiens*), Formosan macaques (*Macaca cyclopis*), common marmosets (*Callithrix jacchus*), golden-backed uakaris (*Cacajao melanocephalus*) and common ravens (*Corvus corax*).

For humans, seven languages are included: Croatian, Greek, Indonesian, US English, Russian, Spanish and Swedish. The data for dolphins come from Ferrer-i-Cancho and Lusseau (2009), for human languages from Ferrer-i-Cancho and Hernández-Fernández (2013), for Formosan macaques from Semple et al. (2010), for common ravens from Conner (1985) and finally, those for common marmosets and golden backed uakaris come from Bezerra et al. (2011).

The mean duration of a call type ($e$ in our notation) is defined as $e = D/f$ where $f$ is the number of times that the call has been produced and $D$ is total duration ($D$ is the sum of all the durations of that call). For common marmosets, $D$ defines two partitions of the repertoire: the low $D$ partition, where the law of brevity is found, and the high $D$ partition, where the law is not found (further details on this partitioning are given in Ferrer-i-Cancho & Hernández-Fernández, 2013). For dolphins, the definition of the size of a behavioral pattern in terms of elementary behavioral units - see Ferrer-i-Cancho & Lusseau, (2009) for a description of these elementary units - is subject to some degree of arbitrariness. For this reason, two variants of behavioral pattern size are considered: one where the elementary behavioral unit "two" is not used and another where elementary unit "stationary" is not used. A summary of the features of the dataset is provided in Table 1.

Table 1 summarizes the results of the randomization test in all the species where the law of brevity has been studied so far with the exception of chick-a-dee calls (Hailman et al., 1985), for which data are not available for reanalysis. The conclusions reached by Hailman et al. (1985) on chick-a-dees, namely that the law of brevity holds in bouts of calls but does not (at least not as clearly) in individual calls must be interpreted carefully. Hailman *et al.*'s (1985) analysis relies on visual inspection of data, which is highly subjective: visual conformity with the law of brevity is clear for bouts of calls but disagreement with the law in individual calls is not obvious. Hailman et al. (1985) did not perform a correlation test to test for conformity to the law of brevity; by contrast, we are determining if the law holds by means of a correlation test between frequency and length/duration. Even raw plots of frequency versus length in human language show substantial dispersion, as shown for example in Fig. 1(a) of Ferrer-i-Cancho & Lusseau (2009).

Table 1. Summary of the features of the dataset and the results of the compression test. Units: s stands for seconds; char stands for characters; e.b.u. stands for elementary behavioral units. $E_{CL}$ was rounded to leave two significant digits. When not bounded, estimated left and right p-values were rounded to leave only one significant digit. Conformity to the law of brevity means that the correlation between frequency and size/duration is negative and significant.

| Species | Behavior | Conformity to law of brevity | $n$ | $E_{CL}$ | left p-value | right p-value |
|---|---|---|---|---|---|---|
| Golden-backed uakaris | Vocalizations | No (Bezerra et al., 2011) | 7 | 0.14 s | 0.09 | 0.9 |
| Common marmosets | Vocalizations | No (Bezerra et al., 2011) | 12 | 0.46 s | 0.5 | 0.5 |
| | Vocalizations (low $D$ cluster) | Yes (Ferrer-i-Cancho & Hernández, 2013) | 6 | 0.048 s | 0.006 | 1 |
| | Vocalizations (high $D$ cluster) | No (Ferrer-i-Cancho & Hernández, 2013) | 6 | 0.59 s | 0.1 | 0.9 |
| Common ravens | Vocalizations | No (Ferrer-i-Cancho & Hernández, 2013) | 18 | 0.25 s | 0.3 | 0.6 |
| Dolphins | Surface behavioral patterns | Yes (Ferrer-i-Cancho & Lusseau, 2009) | 31 | 1.3 e.b.u. | 0.0002 | 1 |
| | *Id.* without "stationary" | | 31 | 1.2 e.b.u. | 0.001 | 1 |
| | *Id.* without "two" | | 31 | 1.3 e.b.u. | 0.002 | 1 |
| Formosan macaques | Vocalizations | Yes (Semple et al., 2010) | 35 | 0.18 s | 0.008 | 1 |
| Humans | Greek | Yes (Ferrer-i-Cancho & Hernández-Fernández, 2012) | 4203 | 3.9 char | $<10^{-5}$ | $>1-10^{-5}$ |
| | Russian | | 7908 | 4.6 char | $<10^{-5}$ | $>1-10^{-5}$ |
| | Croatian | | 15381 | 3.9 char | $<10^{-5}$ | $>1-10^{-5}$ |
| | Swedish | | 19164 | 3.2 char | $<10^{-5}$ | $>1-10^{-5}$ |
| | US English | | 24101 | 3.8 char | $<10^{-5}$ | $>1-10^{-5}$ |
| | Spanish | | 27478 | 3.9 char | $<10^{-5}$ | $>1-10^{-5}$ |
| | Indonesian | | 30461 | 4.2 char | $<10^{-5}$ | $>1-10^{-5}$ |

4. DISCUSSION

Table 1 presents a number of interesting results. First, $E_{CL}$ was significantly small in all cases where conformity to the law of brevity (a significant negative correlation between frequency and size/length/duration) had been reported. This provides further support for the intimate relationship between the law of brevity and the information theoretic principle of compression. Second, there is no evidence of redundancy maximization, the opposite of compression, in the species that we have analyzed: $E_{CL}$ was never significantly high. This result was unexpected for two reasons: pressure for compression is not the only factor shaping repertoires (Bezerra et al., 2011; Ferrer-i-Cancho & Hernández-Fernández, 2013; Endler, 1993) and pressure for compression can cause a signal to be more sensitive to noise in the environment. Coding theory indicates that redundancy must be added in a controlled fashion to combat transmission errors (Cover & Thomas, 2006, pp. 184). In cases where signals can be mistaken for each other, for example due to noise in the environment, forming words by stringing together subunits (e.g. combining 'phonemes' into 'words') allows a system to increase its capacity to communicate (Plotkin & Nowak, 2000). Therefore, redundancy maximization is a conceivable alternative to compression (redundancy minimization).

A range of phenomena or situations may select against compression in signals. Firstly, as mentioned above, environmental noise may drive signals in the opposite direction to that of code minimization. Elongation of signals is one of a number of adaptations to noise, and this is known as the Lombard effect in human language and the communication systems of other species (Zöllinger & Brumm, 2011; Brumm & Zöllinger, 2011); another major strategy to combat noise is to increase redundancy, in the temporal or spatial organization of signals (Richards & Wiley, 1980; Ay et al., 2007). Secondly, capacity for compression may be constrained in some species more than others, due to biological features of the species (Gillooly & Ophir, 2010) or its environment (Wiley & Richards, 1978). For example, the ability to use low-frequency signals to reduce attenuation and thus maximize transmission success is not always possible for smaller animals, due to allometric constraints that limit the frequency (i.e. pitch) of a signal as a function of an animal's body mass (Fletcher, 2004). The same applies to maximizing transmission success by increasing amplitude (Brumm & Slabbekoorn, 2005): a species with a small body cannot produce a sound with high amplitude (Bennet-Clark, 1998; Fletcher, 2004) because amplitude is limited by body mass (Gillooly & Ophir, 2010). Thirdly, reverberation is one of the fundamental problems posed by the forest environment, and one way to overcome this is to package the overall signal into brief pulses that end prior to the delay time expected for the main first reflection of the pulse (Brown & Sinott, 2006, p.191). A continuous signal of duration $t$ could be converted into a package of duration $t + u$, where $u$ is the total duration of the inter-pulse silences. Finally, in relation to human language, elongation occurs in the context of adults' child-directed speech, which tends to be slower, with syllable lengthening, longer pauses, etc. (Saxton, 2010, p. 81 and references therein).

When considering these processes that may drive an increase in signal size, it is important to note that if elongation concerned all the types of elements equally (by a constant proportionality factor), then the law of brevity would remain. This suggests that the law of brevity can only be violated if elongation concentrates on a subset of the repertoire. Consistent with this idea, common marmoset vocalizations do not conform to the law at the level of the whole repertoire, but do within a subset of the repertoire where none of the long-distance calls - calls highly sensitive to noise in the environment and other signal distorting factors - is found (Ferrer-i-Cancho & Hernández-Fernández, 2013). This finding suggests that analysis of logically selected subsets of the vocal repertoire of ravens and golden-backed uakaris, two species where conformity to the law has not been found, is worthwhile.

If mean length maximization plays a role in the species we have examined, it is not strong enough to surface in a statistical test. It is possible that these species introduce redundancy at other levels: below the type of element level, i.e. within the shape or structure of the element, or above the element level, i.e. in the way sequences of elements are constructed. It is also possible that individuals can achieve a similar goal to redundancy by varying instead call amplitude (e.g., Brumm & Slabbekoorn, 2005). The trade-off between compression (mean length minimization) and the need for successful signal transmission, where new element type formation (e.g. words) by combining elements (e.g. 'phonemes') is a well-known strategy (Plotkin & Nowak, 2000), may explain why a clear manifestation of length maximization is difficult to find at our level of analysis.

The existence of true universals in the sense of exceptionless properties is a matter of current debate in human language (Evans & Levinson, 2009) and animal behavior (Bezerra et al., 2011; Ferrer-i-Cancho & Hernández-Fernández, 2013). We believe it is important to investigate the real scope of statistical patterns of language such as the law of brevity in world languages and animal behavior. The analysis of exceptions to the law of brevity is full of subtle statistical and biological details (Ferrer-i-Cancho & Hernández-Fernández, 2013; Bezerra et al., 2011). However, it is of even greater theoretical importance to investigate which are the universal principles of behavior. They are the arena where unification (universality) and the complex nature of reality, including exceptions to language patterns or peculiar situations where certain language patterns emerge, may reconcile. We are not claiming that the law of brevity is a hallmark of language (as opposed to simpler forms of communication), but an example of constrained or convergent evolution by an abstract principle of compression acting upon communicative and non-communicative behavior and reflecting selection for efficiency of coding. This principle provides an avenue for the optimization of the behavior of a species. The exact nature of the solutions adopted may depend on the chances and needs the species encounters during its evolutionary history (Monod, 1972).

ACKNOWLEDGEMENTS

We are grateful to S. E. Fisher, S. Kirby, Rick Dale and three anonymous reviewers for helpful remarks. This work was supported by the grant *Iniciació i reincorporació a la recerca* from the Universitat Politècnica de Catalunya and the grants BASMATI (TIN2011-27479-C04-03) and OpenMT-2 (TIN2009-14675-C03) from the Spanish Ministry of Science and Innovation.